\documentclass{article}
\usepackage{ijcai24}

\usepackage{times}
\usepackage{soul}
\usepackage{url}
\usepackage[hidelinks]{hyperref}
\usepackage[utf8]{inputenc}
\usepackage[small]{caption}
\usepackage{graphicx}
\usepackage{svg}
\usepackage{amsmath}
\usepackage{amsthm}
\usepackage{amsfonts}
\usepackage{booktabs}
\usepackage{algorithm}
\usepackage{algorithmic}
\usepackage[switch]{lineno}

\usepackage{threeparttable}
\usepackage{xcolor}         % colors

\usepackage{xcolor}

\usepackage{etoolbox}

\makeatletter
\patchcmd{\@makefntext}{\insert\footins\bgroup}{\insert\footins\bgroup\let\@makefnmark\relax}{}{}
\makeatother

\newcommand{\footfirstpage}[1]{
  \begingroup
  \renewcommand\thefootnote{} % 重定义脚注编号为空
  \footnotetext{#1} % 添加脚注文本
  \endgroup
}

\title{CoCoG: Controllable Visual Stimuli Generation based on Human Concept Representations}

\author{
Chen Wei$^{\dagger,1,2}$\and
Jiachen Zou$^{\dagger,1}$\and
Dietmar Heinke$^2$\And
Quanying Liu$^{*,1}$\\
\affiliations
$^1$Southern University of Science and Technology, Shenzhen, China \\
$^2$University of Birmingham, Birmingham, United Kingdom\\
\emails
\{weic3, zoujc2022\}@mail.sustech.edu.cn,
d.g.heinke@bham.ac.uk,
liuqy@sustech.edu.cn
}

\begin{document}

\maketitle

\footfirstpage{$\dagger$ Equal contribution.}
\footfirstpage{$^*$ Corresponding author.}
\footfirstpage{This paper has been accepted by IJCAI2024.}

\begin{abstract}
%Humans construct internal concepts through external visual stimuli. How to uncover human low-dimensional concept representation space and how to generate high-dimensional visual stimuli by controlling concepts are core technical challenges.

A central question for cognitive science is to understand how humans process visual objects, i.e, to uncover human low-dimensional concept representation space from high-dimensional visual stimuli.
Generating visual stimuli with controlling concepts is the key. However, there are currently no generative models in AI to solve this problem. 
Here, we present the Concept based Controllable Generation (CoCoG) framework.
CoCoG consists of two components, a simple yet efficient AI agent for extracting interpretable concept and predicting human decision-making in visual similarity judgment tasks, and a conditional generation model for generating visual stimuli given the concepts. 
We quantify the performance of CoCoG from two aspects, the human behavior prediction accuracy and the controllable generation ability. The experiments with CoCoG indicate that 1) the reliable concept embeddings in CoCoG allows to predict human behavior with 64.07\% accuracy in the THINGS-similarity dataset; 2) CoCoG can generate diverse objects through the control of concepts; 
3) CoCoG can manipulate human similarity judgment behavior by intervening key concepts.
CoCoG offers visual objects with controlling concepts to advance our understanding of causality in human cognition. The code of CoCoG is available at \url{https://github.com/ncclab-sustech/CoCoG}.

\end{abstract}

\section{Introduction}

Humans receive abundant visual stimulation from the natural world. 
Unlike computer vision models that aim at object recognition tasks, humans aim to survive in complex nature, which requires understanding abstract concepts of these visual objects, such as functionality, toxicity, and danger.
To explore such concept representation in humans, scientists have proposed a series of visual stimuli-based decision-making tasks~\cite{murphy1985role,medin1993respects,hebart2020revealing,roads2023modeling}, such as the similarity judgment task, in which visual stimuli of specific concepts are presented to the participants and then their decision-making behaviors are recorded. 
However, understanding human concept representation through these cognitive tasks has two main difficulties. On the one hand, training an AI agent to uncover interpretable concept representations by predicting human decision-making behavior requires substantial human decision-making data under massive visual objects. On the other hand, to understand the causal relationship between concept representation and behavior, it is necessary to manipulate the concepts, preserving all other low-level features, to generate visual objects. This is unexplored territory for AI.
It poses a new technical challenge for image generation, namely \textit{controllable visual object generation based on concept representation}.

\begin{figure}[t]
\includegraphics[width=9cm]{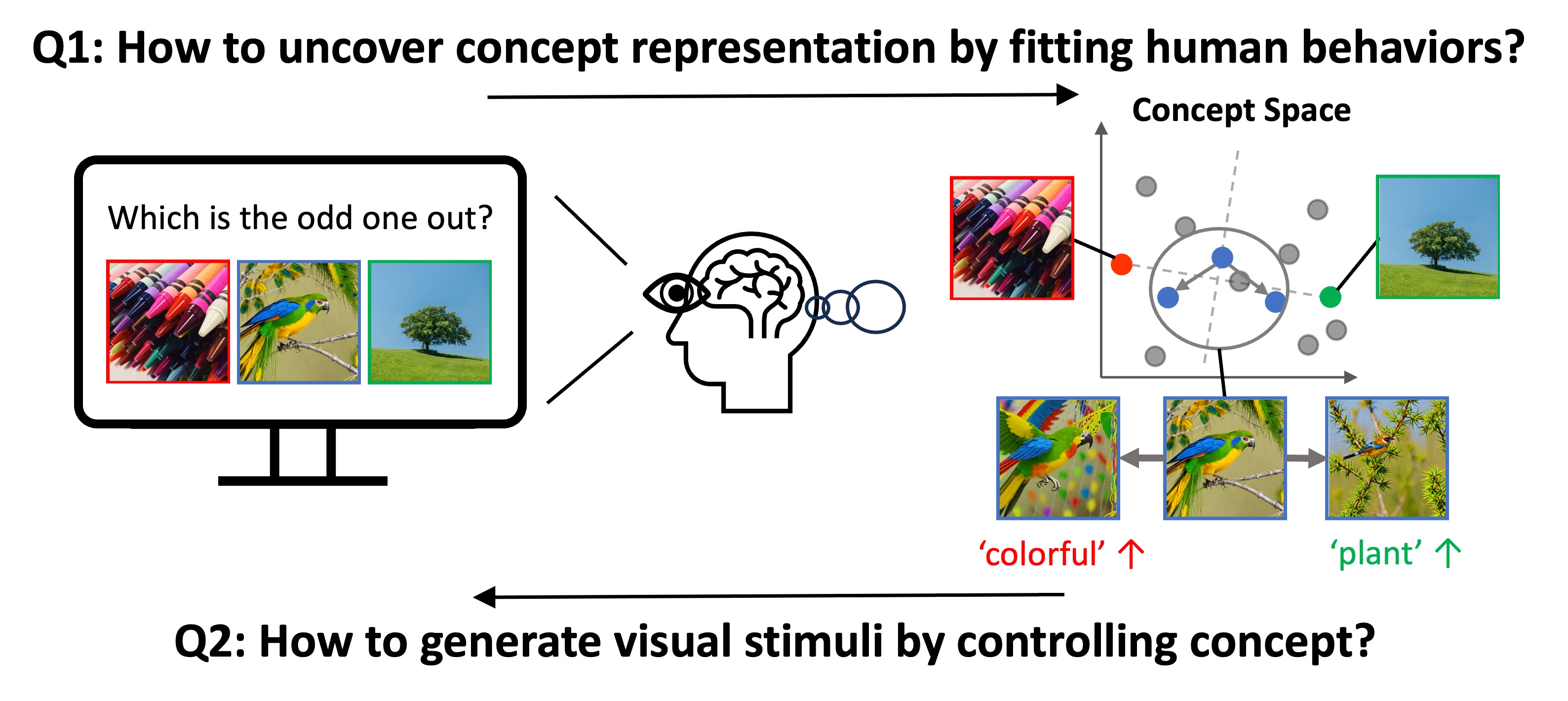}
\caption{Motivations of our work. %1) Uncovering the concept representation from behavior; 2) Controllable generation to manipulate human behavior.
}
\label{fig:framework}
\end{figure}

Controllable generation models have made great progress in AI community, for example, conditional generation models based on GAN~\cite{goodfellow2020generative,tao2022df} and diffusion~\cite{song2020score,ho2020denoising,ho2022classifier}. These conditional generation models have been applied in many fields, including text-to-image/video generation~\cite{rombach2022high,ramesh2022hierarchical}, image inverse problem~\cite{kawar2022denoising,chung2022improving,meng2021sdedit} and biomedical imaging\cite{song2021solving,ozbey2023unsupervised}. The conditions for controllable image generation can come from multiple aspects, such as text, sketches, edge maps, segmentation maps, depth maps~\cite{rombach2022high,ramesh2022hierarchical,meng2021sdedit,zhang2023adding,yu2023freedom,bansal2023universal}. 
However, these conditions do not include human subjective feelings nor human feedback, resulting in generated images misaligned with human needs. For instance, generated images by recommendation system may not meet human preference.
To align the generated images with human needs, some pioneer works brought human feedback (e.g., the human visual preference score obtained offline ~\cite{wu2023human,kirstain2023pick}, the human-in-the-loop visual comparison decision~\cite{von2023fabric,fan2023dpok,tang2023zeroth}) into conditional generative models.
Nevertheless, these works have not consider prior knowledge of cognitive science, such as the factors that have the greatest impact on human decision-making, that is, concepts, as control variables for image generation.

A large number of human research suggest that similarity judgment tasks are an effective experimental paradigm for revealing human concept representations~\cite{roads2023modeling,hebart2020revealing}. In these tasks, human subjects are asked to compare different visual stimuli and make decisions based on their similarity. AI models, such as~\cite{peterson2018evaluating,marjieh2022words,marjieh2022predicting,muttenthaler2022human,jha2023extracting,fu2023dreamsim}, are proposed to predict the subjects' decisions, under an assumption that human perception of visual objects can be encoded into a low-dimensional mental representation, namely \textit{concept embedding}.
The distance between visual objects in this concept space reflects the distance of visual objects in the human's mind. Human decision-making behavior can be explained by different dimensions of concept representation. Thus, aligning AI models with humans in terms of concept representation would naturally align their outputs as well.
Also, compared with existing controllable image generation methods, a controllable generation model based on concept representation would control human decision-making behavior more effectively.

In this study, we propose a Concept based Controllable Generation (CoCoG) framework. CoCoG utilize concept embeddings as conditions for the image generation model, bridging cognitive science and AI. %Specifically, for cognitive science, it efficiently and controllably generates natural visual stimuli, and for visual generative models, it makes the model's performance more aligned with human cognition.
CoCoG comprises two parts: a \textit{concept encoder} for learning concept embedding via predicting human behaviors and a \textit{concept decoder} which employs a conditional diffusion model for mapping concept embedding to visual stimuli through a two-stage generation strategy.
%For the concept encoder, we mimic human visual perception and use DNNs to learn the mapping from visual objects to concept embeddings. For the concept decoder, we employ a conditional diffusion model and build a mapping from concept embeddings to visual objects through a two-stage generation strategy. 

CoCoG has three main contributions:
\begin{itemize}
    \item CoCoG's concept encoder can predict human visual similarity decision-making behaviors with higher accuracy than the state-of-the-art (SOTA) model (i.e., VICE~\cite{muttenthaler2022vice}), uncovering a reliable, interpretable concept space of humans (Figure~\ref{fig:concept encoder}).
    \item CoCoG's concept decoder can generate visual stimuli by controlling the concept embedding. The generated visual stimuli have diversity and high consistency with the target concept embeddings (Figure~\ref{fig:generated images}\&\ref{fig:distribution and guidance scale}). The generation can be guided with text prompts (Figure~\ref{fig:multi_prompt}\&\ref{fig:multi_concept}).
    \item CoCoG can manipulate the human similarity decision-making behavior in a highly controllable way by controlling the concepts of generated visual stimuli (Figure~\ref{fig:intervene}).
\end{itemize}

%%%%%%%%%%%%

\begin{figure*}[!h]
\centering
\includegraphics[width=15cm]{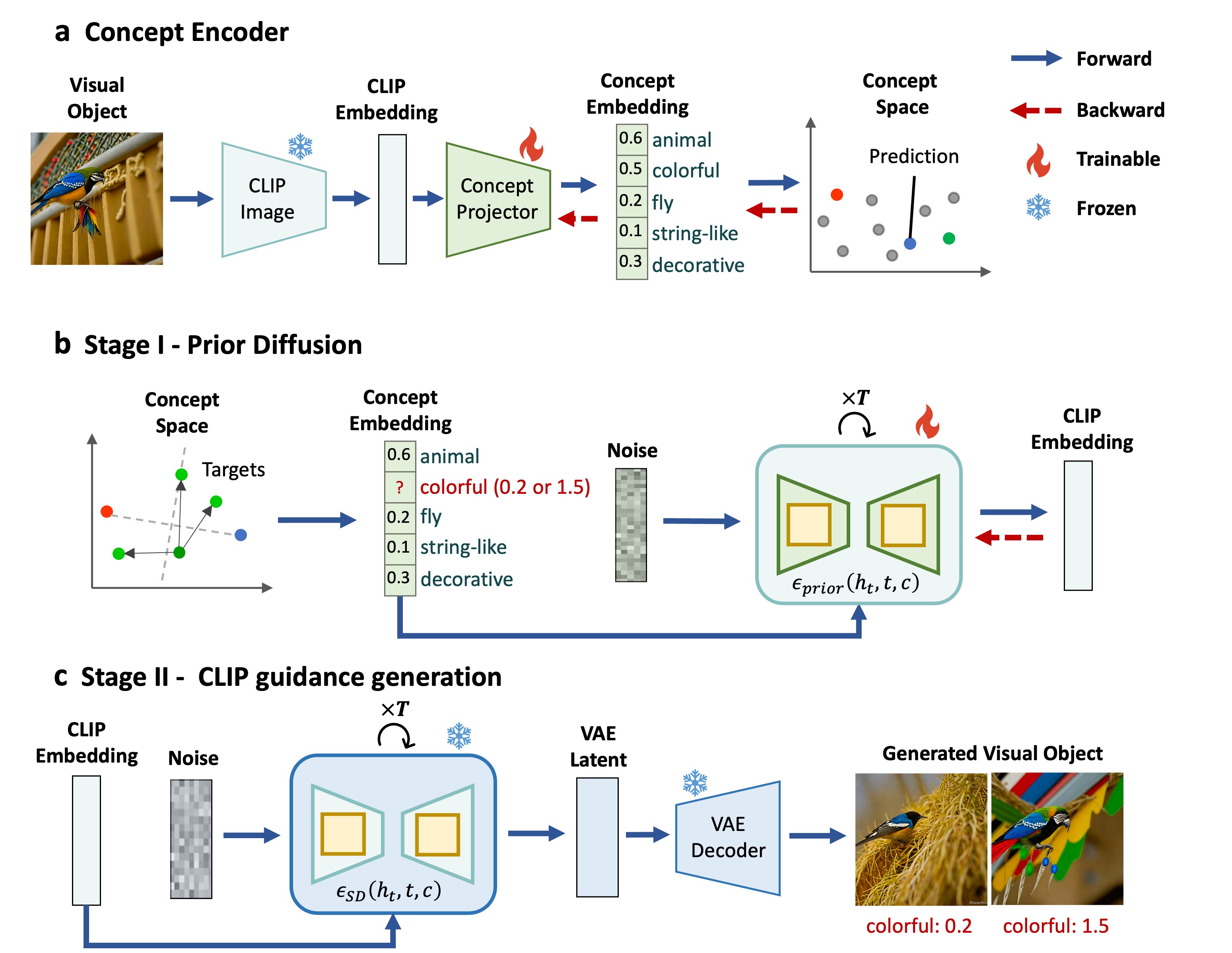}
\caption{The framework of CoCoG. (a) The \textit{concept encoder} for learning concept embeddings using a similarity judgment behavior dataset. Visual objects are processed through the CLIP image encoder to obtain CLIP image embeddings, and then passed through a learnable concept projector to obtain concept embeddings. Then, we can predict similarity judgment behaviors by compute similarity with others. (b) The stage I of the concept decoder, the \textit{prior diffusion} for determining the concept embedding based on our desired judgment behavior (e.g., here modifying the concept “colorful”). Then, we train a diffusion model conditioned on the concept embedding to generate the corresponding CLIP embedding. (c) Stage II of the concept decoder, the \textit{CLIP guided generation}. It uses the CLIP embedding as a condition to guide the pre-trained image diffusion generation model to generate VAE latent, which are then processed through the VAE decoder to produce the generated visual object. }
\label{fig:framework}
\end{figure*}

\section{Method}

The CoCoG method comprises two parts: a concept encoder and a two-stage concept decoder, as shown in Figure \ref{fig:framework}.

%As shown in Figure \ref{fig:framework}, our method comprises two parts: the concept encoder and the concept decoder.

\subsection{Concept encoder for embedding low-dimensional concepts}

The first step in CoCcoG is to train a concept encoder to learn the concept embeddings of visual objects (Figure~\ref{fig:framework}a). Given a dataset of visual objects $X$, for each visual object $x$ in the dataset, we first input it into the CLIP image encoder $f$ to extract the CLIP embedding $h \in \mathcal{R}^{D}$. Then, the CLIP embedding is input into a learnable concept projector $g$, thereby generating the concept embedding of the visual object $c \in \mathcal{R}^{d}$:
\begin{equation}
\begin{aligned}
&\text{CLIP embedding:} & h = f(x),\\
&\text{concept embedding:} & c = g(h),
\end{aligned}
\end{equation}
where the each dimension in concept embedding $c$ represents aninterpretable concept, and the corresponding activation of this dimension indicates the activation strength of the concept in the visual object. In short, we have $ c = g(f(x)) $.

To train this model, we used the {\it triplet odd-one-out} similarity judgment task in the THINGS dataset~\cite{hebart2023things}. In this task, participants are asked to view three visual objects and consider the similarity between each pair of visual objects to determine the similar pair and select the remaining one as the {\it odd-one-out}.

Similar to previous works~\cite{zheng2018revealing,hebart2020revealing,muttenthaler2022vice}, we use the dot product similarity as the similarity measurement function between concept embeddings (i.e., $S_{ij} = <c_i, c_j>$) and use cross-entropy based on pairwise similarities to predict human decisions. Therefore, for the concept embeddings $c_i, c_j, c_k$ of three visual objects $x_i, x_j, x_k$, we have
\begin{equation}
p(y) = CrossEntropy(S_{jk}, S_{ik}, S_{ij}),
\label{eq:similarity}
\end{equation}
where $p(y)$ is the probability distribution of the triplet visual stimuli $(x_i, x_j, x_k)$ being the {\it odd-one-out}, with the highest probability choice being the model's predicted behavioral outcome.
By comparing the model's predicted behavioral outcomes with the recorded behavioral outcomes from human participants, we calculate the loss and perform backpropagation to train the concept projector $g$. In the specific process, we added an $L_1$ regularization to $c$ to ensure the sparseness of low-dimensional concept embedding. We verified different training strategies (see Appendix).

\subsection{Two-stage concept decoder for controllable visual stimuli generation}

After training the concept encoder, we can obtain the data triplet $(x_i, h_i, c_i)$ for each image in the dataset. Based on these data, we then train a concept decoder to generate visual objects by controlling human conceptual representations. 

In our model, the concept embedding $c$ completely determines the distribution of the human decision $y$, and the $h$ completely determines the distribution of $c$. Therefore, the joint distribution of the human decisions $p(x, h, c, y)$ can be formulated as:
\begin{equation}
p(x, h, c, y) = p(y)p(c|y)p(h|c)p(x|h).
\label{eq:2-stage}
\end{equation}

Next, we present each step of $p(c|y), p(h|c), p(x|h)$ respectively.
$p(c|y)$ means finding the appropriate concept embedding $c$ given the behavioral outcome $y$. This step is determined according to specific control objectives, which we will discuss in chapter~\ref{ch:control objectives}.
For $p(c|y)$ and $p(x|h)$, we decompose the concept decoder into two stages, the Prior Diffusion and the CLIP guided generation, respectively, which execute the processes of generating CLIP embedding $h$ from the concept embedding $c$ and generating visual object $x$ from the CLIP embedding $h$.

\paragraph{Stage I - Prior diffusion}

Inspired by DALL-E 2~\cite{ramesh2022hierarchical}, we train a diffusion model conditioned on the concept embedding $c$ to learn the distribution of CLIP embeddings $p(h|c)$ (Figure~\ref{fig:framework}b). The CLIP embedding $h$ obtained in this stage serves as the prior for the next stage. We construct a lightweight U-Net: $\epsilon_{prior}(h_t, t, c)$, where $h_t$ represents the noisy CLIP embedding in the diffusion time step $t$. We extract the CLIP embeddings and the concept embeddings from ImageNet and use the obtained training pairs $(h_i, c_i)$ to train the Prior Diffusion model. We employ the classifier-free guidance method~\ to train this conditional generative diffusion model. The specific formulas can be found in the Appendix.

\paragraph{Stage II - CLIP guidance generation}

After obtaining the CLIP embedding $h$ in Stage I, we model a generator $p(x|h)$ to sample the visual object $x$ conditional to $h$ (Figure~\ref{fig:framework}c). In this study, we use the pre-trained SDXL and IP-Adapter models\cite{podell2023sdxl,sauer2023adversarial,ye2023ip}. Speicfically, SDXL serves as the backbone of the image diffusion generation model. Through the dual cross-attention modules of the IP-Adapter, we input the CLIP embedding $h$ as a condition, thereby guiding the denoising process of the U-Net. Similar to Stage I, we have the model $\epsilon_{SD}(z_t, t, h)$, where $z_t$ are the noisy latents of SDXL's VAE. Details can be found in the Appendix. Since we freeze the pre-trained models without any modification, we can simply use the existing functionalities of the pre-trained models and combine them with concept embedding guidance.

\begin{figure}[!h]
\centering
\includegraphics[width=7.5cm]{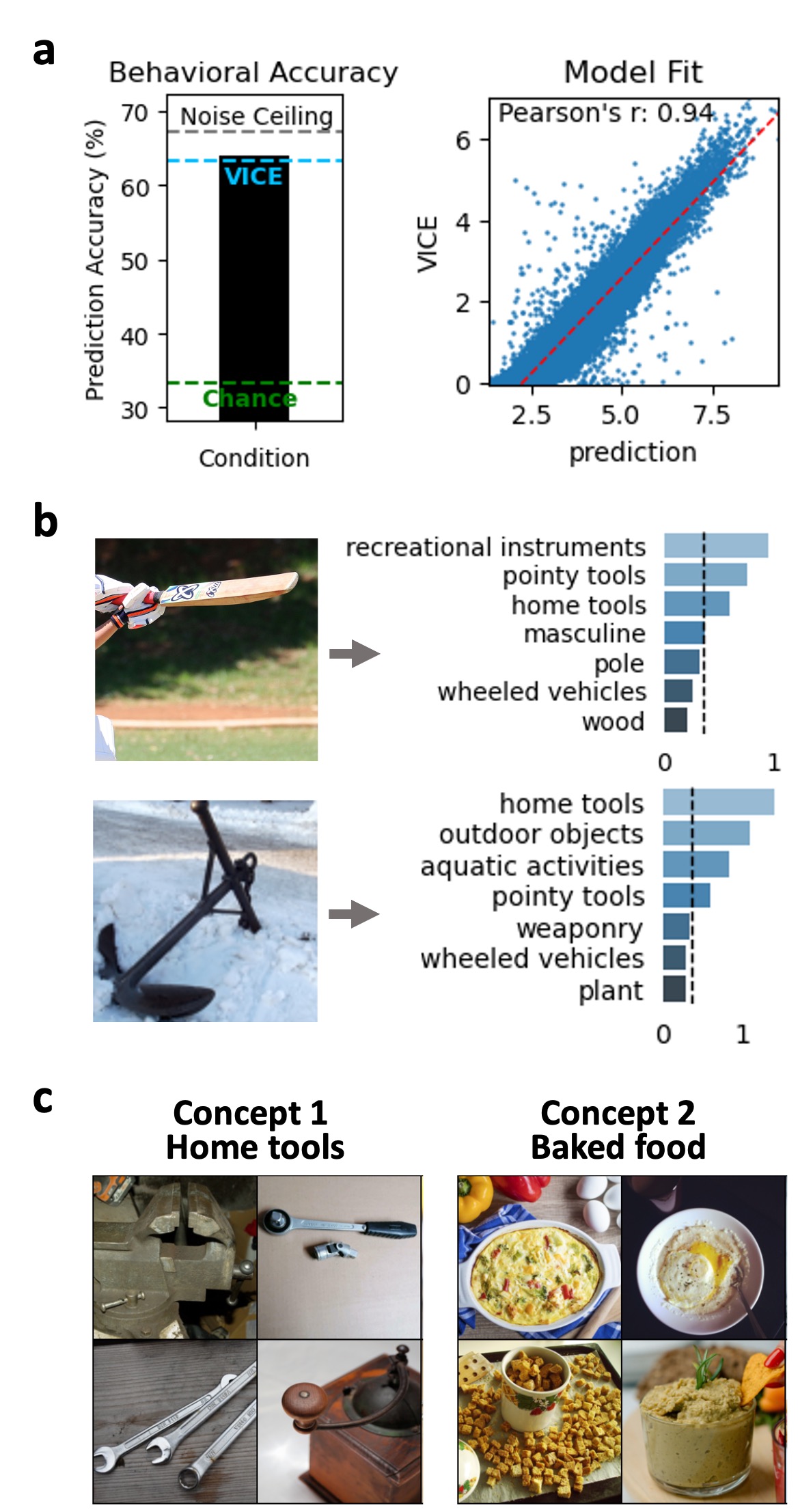}
\caption{The performance of the concept encoder in predicting and explaining human behavior. (a) Our model's prediction accuracy for similarity judgment behavior is 64.07\%, exceeding the previous SOTA model VICE's 63.27\% (blue dashed line), with only slightly lower than  the noise ceiling (gray dashed line)%~\cite{hebart2020revealing}. 
The Pearson correlation coefficient between the similarity of visual objects predicted by our model and by VICE is 0.94; (b) Example visual objects and their concept embeddings, with dashed lines representing the 90th percentile of activated concepts; (c) Example visual objects with significant activation on the concept \textit{Home tools} and \textit{Baked food}, respectively.}

\label{fig:concept encoder}
\end{figure}

\paragraph{CLIP embedding as an intermediate variable} In both processes from visual objects to concept embeddings and from concept embeddings back to visual objects, we use CLIP embeddings as an intermediate variable. This is for two purposes:
1) For the concept encoder, CLIP embeddings are sufficiently low-dimensional and contain key information of images. Previous study has shown that using CLIP embeddings along with simple linear probing can well predict human behavior in similarity judgment tasks~\cite{muttenthaler2022human}.
2) For the concept decoder, using CLIP embeddings helps us better utilize existing pre-trained conditional generative models. Existing models can already use CLIP embeddings as a conditional input for the generative model, allowing us to simply adopt a two-stage generation strategy. We only need to train a Prior Diffusion model, which significantly reduces the computational cost of training and inference.

    %%%%%%%%%%%%
\section{Model Validation}

\subsection{Concept encoder can predict and explain human behaviors}

We first validated the concept encoder from two aspects: the prediction accuracy of human behaviors (Figure~\ref{fig:concept encoder}a) and the interpretability of the learned lwo-dimensional concepts (Figure~\ref{fig:concept encoder}b\&c). We used the THINGS Odd-one-out dataset as the similarity judgment behavior dataset to train our concept encoding model. We tested various configurations of the concept encoding model (which will be detailed in the Appendix).  Figure~\ref{fig:concept encoder}a shows the results of the optimal model configuration. We used 42-dimensional concept embeddings for comparison with previous SOTA model (VICE) and our experiments have found that more than 42 dimensions only bring small improvements.
% Do we need to discuss the number of dimensions in detail here as mentioned in the rebuttal? It feels like it’s unnecessary.
In terms of behavioral prediction, our best-performing model achieved an accuracy of 64.07\% on the THINGS Odd-one-out dataset, surpassing the previous best model VICE's accuracy of 63.27\%. 
% Do we need to discuss generalization performance in detail here as mentioned in the rebuttal? It feels like it's missing the point a bit.
We also compared the similarity predictions of visual objects between our model and VICE, and the Pearson correlation coefficient of their predictions reached 0.94. This indicates that our model can accurately predict human similarity judgment behaviors.

One major merit of our concept embedding is its good interpretability. Figure~\ref{fig:concept encoder}b shows the low-dimensional concept embeddings for the two example visual objects. It is obvious that the concept embeddings encoded by CoCoG well describe the visual objects, and the activation of concepts in a visual object exhibits a favorable property of sparsity, which is in line with human intuition (see Appendix).
We used the CLIP-Dissect model to automatically choose words from the lexicon to describe the concepts in each of the 42 dimensions. Note that the relationship between these words and the concepts is not absolute and is only for examples.
Figure~\ref{fig:concept encoder} shows the visual objects with the highest activation in the first two conceptual dimensions (i.e., home tools and baked food). Importantly, we find that the visual objects that significantly activate these dimensions are highly consistent with these two concepts. This indicates that our model can effectively encode the conceptual embeddings of visual objects, and the encoded concepts have good interpretability.

\subsection{Concept decoder can generate visual objects consistent with concept embedding}

\begin{figure}[!ht]
\centering
\includegraphics[width=7.5cm]{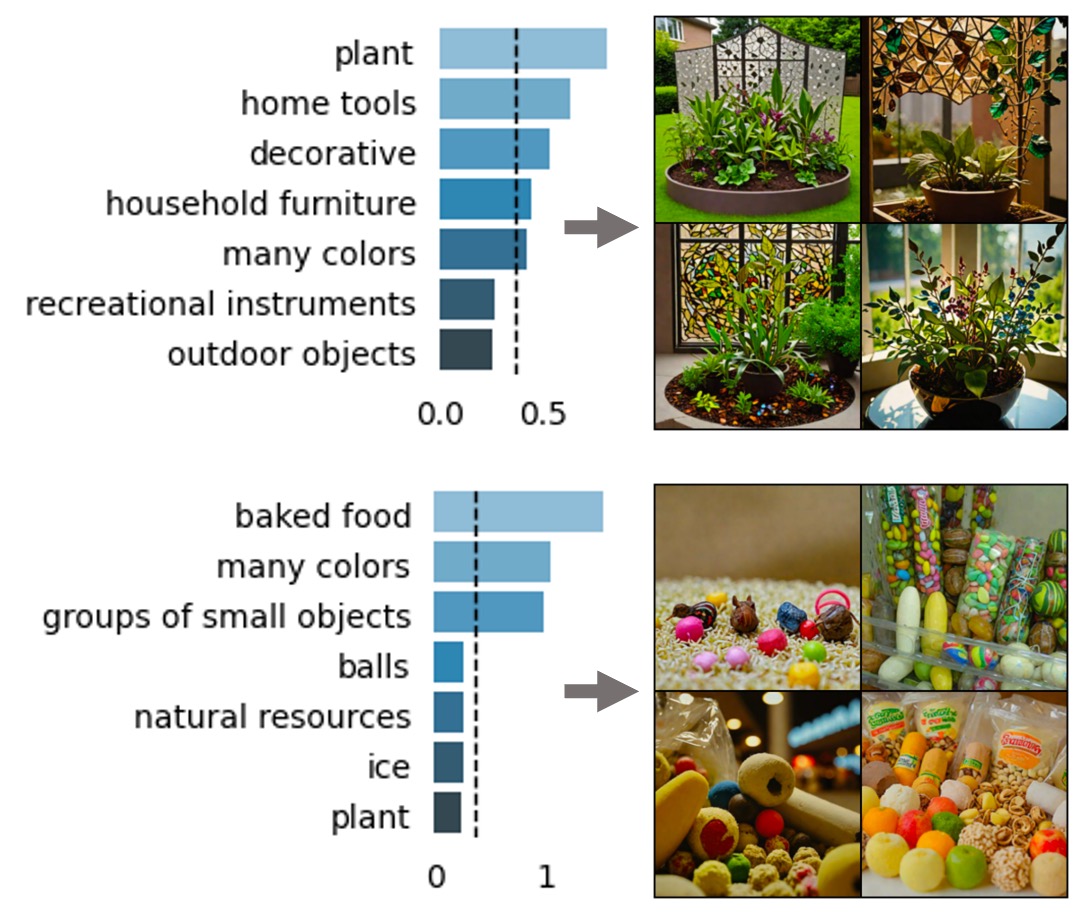}
\caption{The visual objects generated by controlling the concept embeddings.}
\label{fig:generated images}
\end{figure}

In this section, we validate the generative effectiveness of the concept decoder. Specific training parameters are shown in the Appendix.
Figure~\ref{fig:generated images} shows visual objects generated under the guidance of concept embeddings. %The figure presents visual objects generated through the concept decoder from concept embeddings. 
These visual objects generated from the same concept embedding are well-aligned with the concept embedding and have good diversity, demonstrating that our diffusion model can conditionally generate visual objects consistent with the concept embeddings.

\begin{figure}[!ht]
\centering
\includegraphics[width=7.5cm]{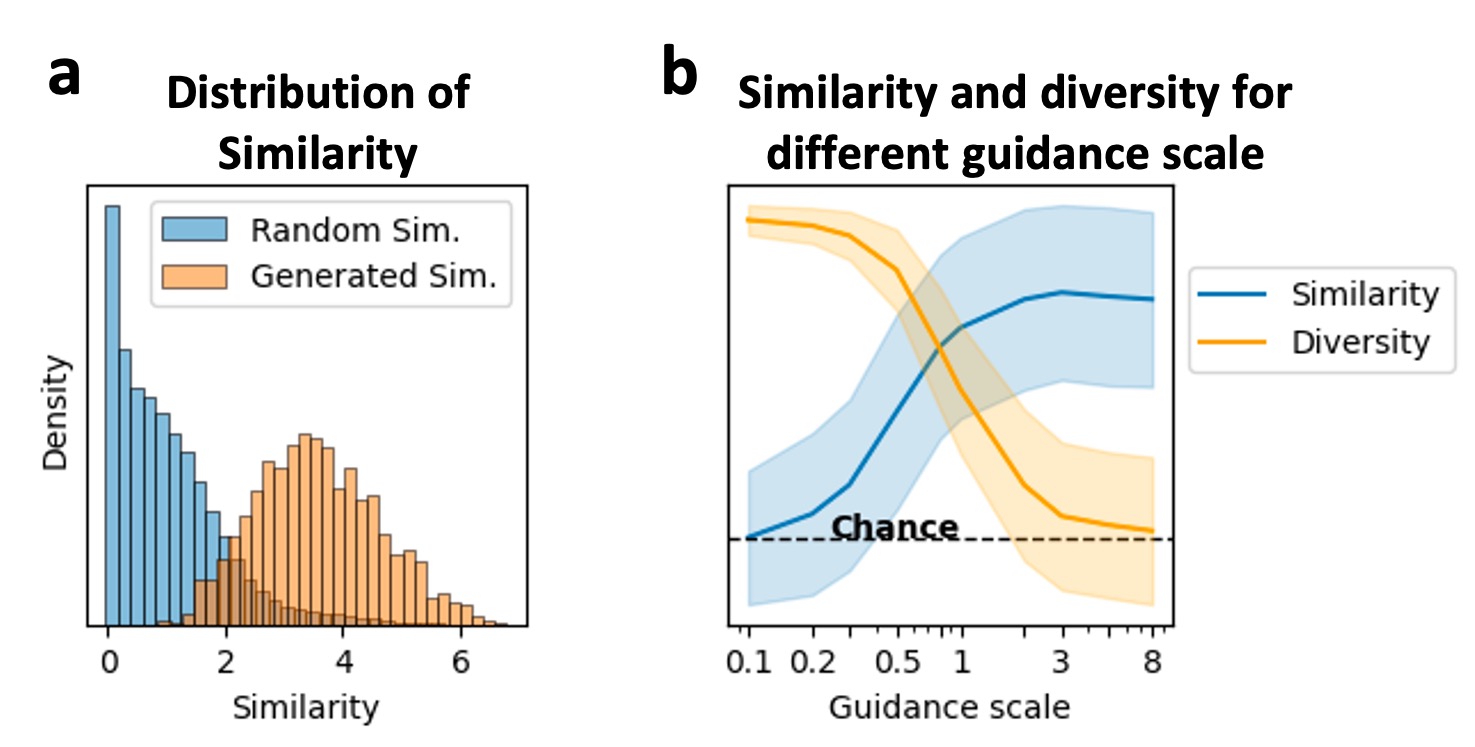}
\caption{Measurements of the performance of generated visual objects. (a) The similarity between random visual objects and the target concept embeddings (blue), and the similarity between generated visual objects and the target concept embeddings (orange); (b) The similarity between visual objects and target concept embeddings as the guidance scale changes (blue), and the diversity of visual objects as the guidance scale changes (orange).}
\label{fig:distribution and guidance scale}
\end{figure}

We quantify the similarity between the concept embeddings of the generated visual objects and the target concept embeddings according to Eq.\ref{eq:similarity}. The similarity between generated visual objects and target concept embeddings is significantly higher than that between two random visual objects (Figure~\ref{fig:distribution and guidance scale}a). Additionally, by adjusting the guidance scale, we can control the guiding strength of the target concept embeddings, thereby controlling the similarity and diversity of the generated visual objects (specific metrics calculation can be found in the Appendix). As the guidance scale increases, the similarity between the visual objects and the target concept embeddings increases while the diversity decreases (Figure~\ref{fig:distribution and guidance scale}b).

\section{CoCoG for studying counterfactual explanations of human behaviors}

According to the role of concept embedding, manipulating the concept embedding can directly influence human similarity judgment behavior. Conversely, does the change in visual objects that do not affect the concept embedding have no impact on human behavior? CoCoG is an excellent tool to explore this counterfactual question. 

\subsection{Flexible controlling of generated objects with text prompts}

\begin{figure}[t]
\centering
\includegraphics[width=7.5cm]{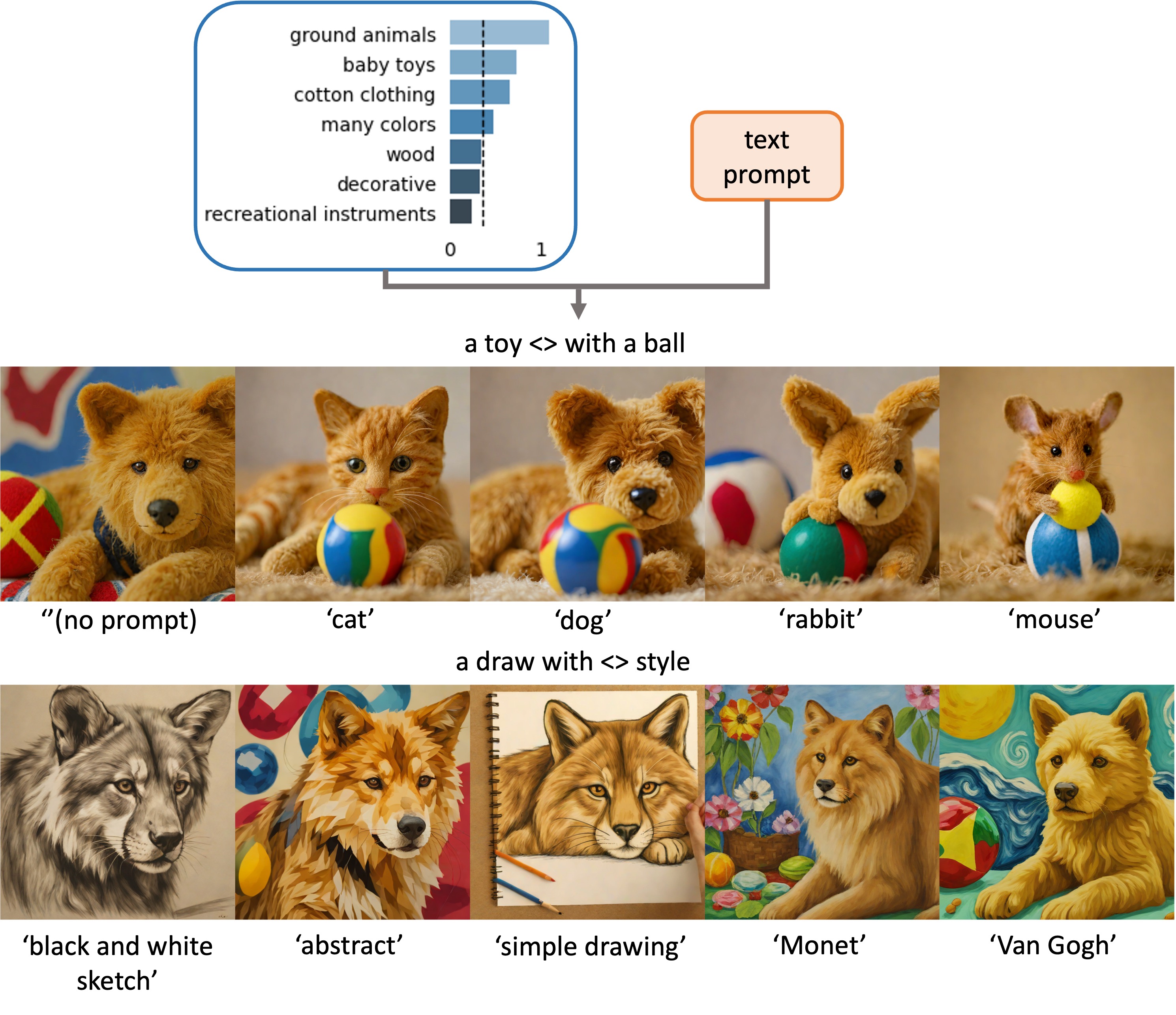}
\caption{Visual objects generated with the same concept embedding combined with different text prompts. The first image in the first row is generated without using any text prompt, while other images are generated by using different animal species as text prompts. The images in the second row are generated by using different artistic styles as text prompts.}
\label{fig:multi_prompt}
\end{figure}

As shown in \ref{fig:multi_prompt}, we used the same concept embedding combined with different text prompts to generate visual objects. Regardless of the changes in the category and style of the visual objects, the images consistently retained the characteristics of the concept embedding. Note that we fixed the random seed to highlight the differences in text prompts. In theory, these completely different visual objects would lead to similar judgment behaviors in similarity judgment experiments.

\begin{figure}[!ht]
\centering
\includegraphics[width=7.5cm]{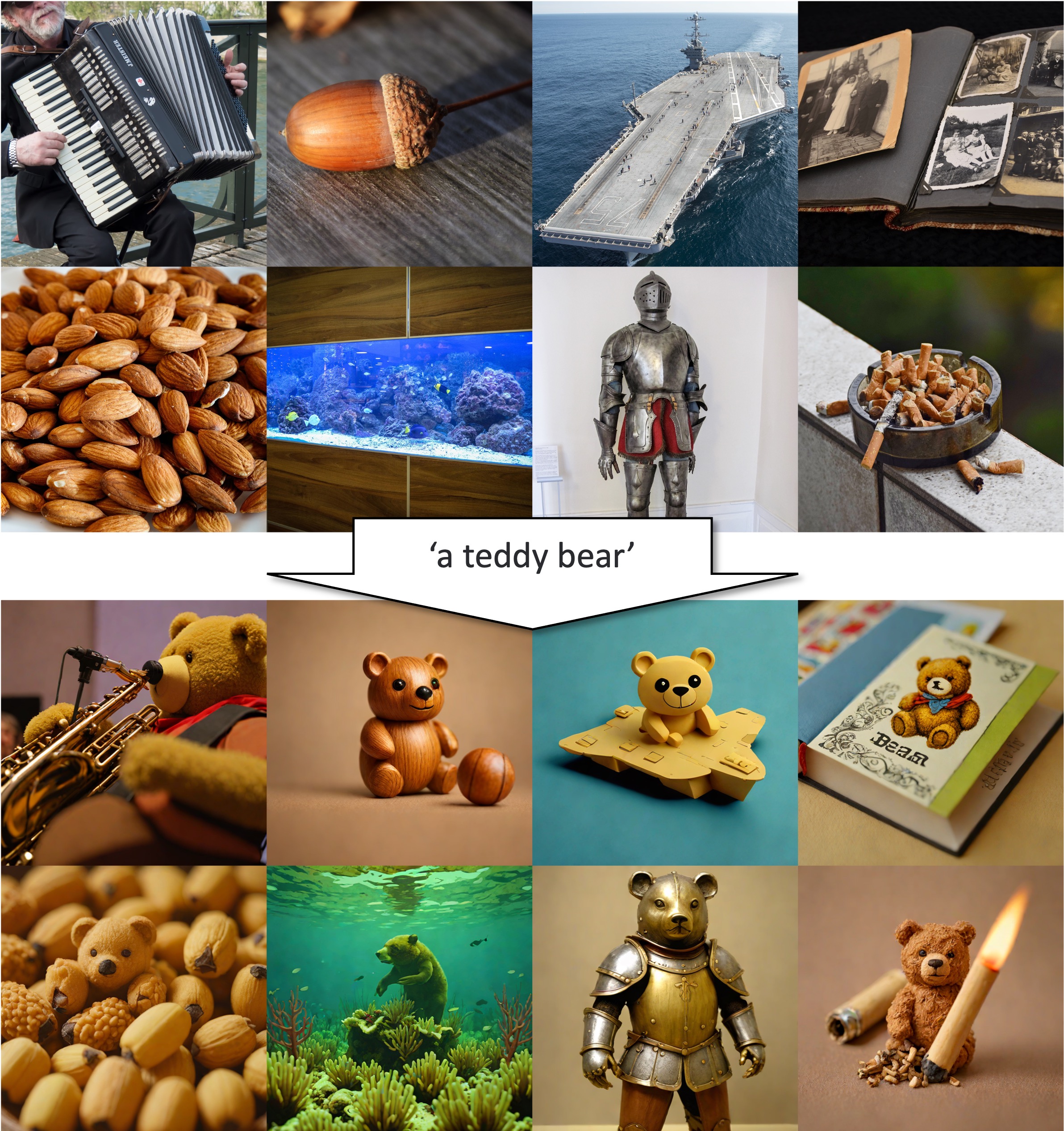}
\caption{Visual objects generated with different concept embeddings combined with the same text prompt. The images in the upper and lower halves correspond to each other. The images in the upper half were used to extract concept embeddings, which were then combined with the text prompt `a teddy bear' to generate the new images in the lower half.}
\label{fig:multi_concept}
\end{figure}

\begin{figure*}[!h]
\centering
\includegraphics[width=15cm]{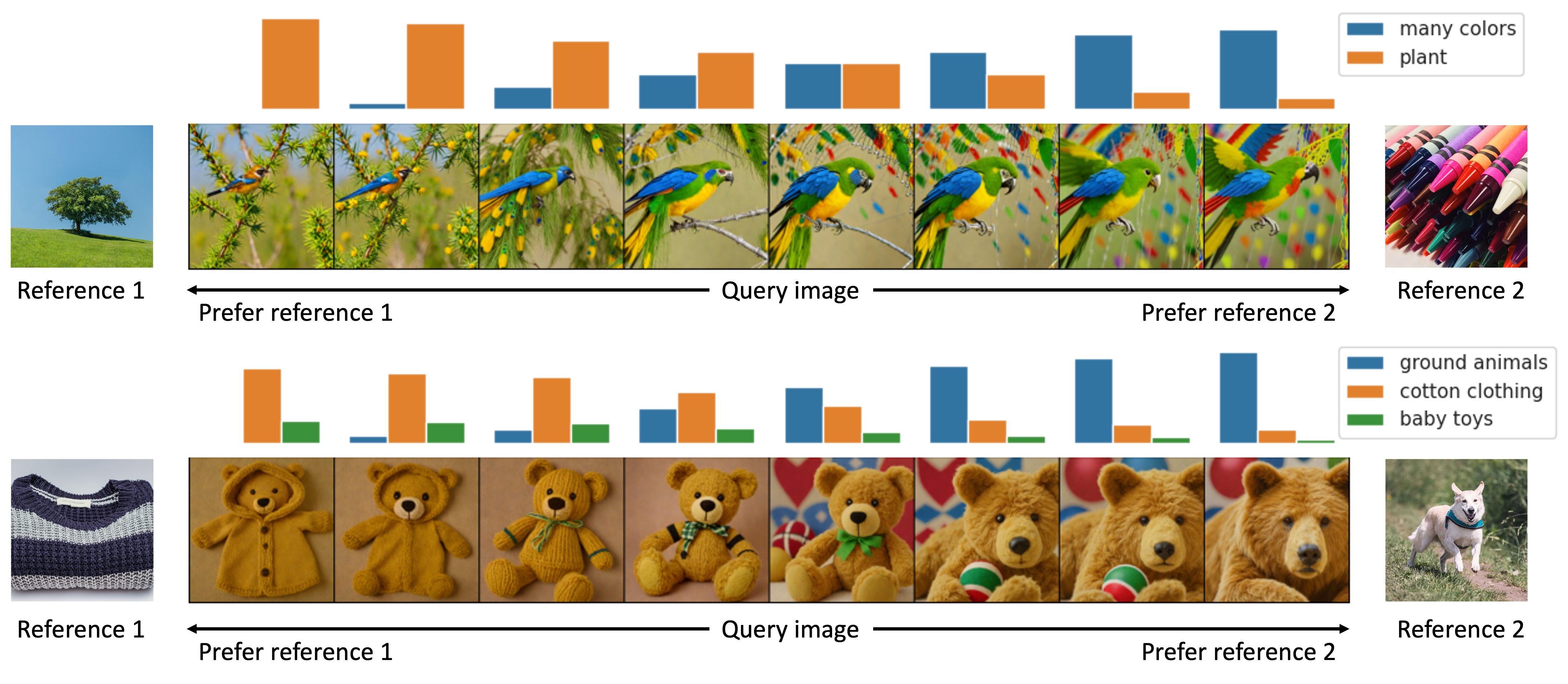}
\caption{Manipulating similarity judgment behavior by key concepts intervention. (upper) The visual objects generated by using ``many colors" and ``plant" as key concepts. (bottom) The visual objects generated by using ``ground animals", ``cotton clothing", and ``baby toys" as key concepts.}
\label{fig:intervene}
\end{figure*}

Further, we show visual objects generated using the same text prompt combined with different concept embeddings (Figure~\ref{fig:multi_concept}). The images in the upper half were used to extract concept embeddings, which were then combined with text prompts to generate the new images in the lower half. The generated images reflect the property "teddy bear" after adding the text prompt, with preserving the concept embeddings. Therefore, contrary to the concept-editing experiment, modifying text prompts do not lead to changes in judgment behavior because the concept embedding is preserved.

%%%%%%%%%%%%
\subsection{Manipulating the similarity judgment decisions by intervening the key concepts}
\label{ch:control objectives}

In this experiment, we designed a simple scenario to show how CoCoG can manipulate similarity judgment behavior. Suppose participants need to perform a {\it Two alternative forced choice} experiment, i.e., they need to choose from Reference 1 \& 2 which is more similar to the Query. In this scenario, we can directly increase or decrease certain concepts to make the Query more similar to one of the references. Considering that we use dot product similarity, we can simply choose the concept that is significant in the Reference but not in the Query as the key concept. The results in Figure~\ref{fig:intervene}, show that the preference of the Query for Reference 1 \& 2 is manipulated by modifying two and three concepts, with the concepts shifting to two different directions. It is evident that the generated visual objects by CoCoG can change smoothly but significantly with the manipulation of concepts.

This method provides us with an effective tool for analyzing the causal mechanisms of concepts in similarity judgment tasks. For example, in the experiment, we can directly manipulate the concepts of interest and observe subsequent behavior. Researchers can actively choose to explore the impact of certain concepts on behavior and analyze the causal relationship between changes in concepts and judgment behavior. By focusing on informative concepts and precisely controlling the activation values of concepts, this method is expected to improve the efficiency of data collection and significantly reduce the number of experimental trials needed (rather than using millions of trials~\cite{hebart2023things}).

%%%%%%%%%%%%
\section{Related Works}

% \subsection{...}

% \paragraph{``...it would have been useful to have a better positioning of the current contribution w.r.t. prior work (especially those using the CLIP neural network as well).''}
% As you pointed out, many studies have used DNNs, including CLIP, to model human similarity judgment tasks, which we have reviewed in Section 5.1. Section 2.2's final paragraph explains why we use CLIP embeddings as an intermediate variable.
% In contrast, our work focuses on generating visual stimuli from given concepts. This allows our model to have more potential applications beyond modeling human similarity judgment behavior. 
% As you said, we will emphasize this point more clearly if accepted.
% I don't know how to change it www

\subsection{Concept Embeddings}

Numerous computational models have been developed to model embeddings encoding in similarity judgment tasks~\cite{roads2023modeling,hebart2020revealing}. These methods aim to study concept embeddings through similarity judgment tasks, either by optimizing concept embeddings for each object~\cite{zheng2018revealing,roads2021enriching,hebart2020revealing,muttenthaler2022vice} or by using DNNs to predict human similarity judgment behavior~\cite{peterson2018evaluating,marjieh2022words,marjieh2022predicting,muttenthaler2022human,jha2023extracting,fu2023dreamsim}. 
Previous methods, assuming sparsity, continuity, and positivity of concept embeddings, learn low-dimensional concept embeddings for each object through probabilistic models. These embeddings have good interpretability but cannot generalize to new objects. On the other hand, DNN-based methods construct concept embeddings from neural activations by reducing the dimensionality of high-dimensional DNN latent representations~\cite{jha2023extracting}, aligning DNN-based similarity judgments with human judgments~\cite{muttenthaler2023improving}, or using multimodal inputs to improve similarity judgment predictions~\cite{marjieh2022words,marjieh2022predicting}. This approach can utilize state-of-the-art DNN models to form embeddings that naturally generalize to new objects. 
To enhance behavioral experiments, Roads et al. utilized active learning coupled with a trial selection strategy for efficient concept embedding inference~\cite{roads2021enriching}. Similarly, DreamSim leveraging Stable Diffusion, generates synthetic data within specified categories to create a Perceptual dataset, aimed at studying Human perceptual judgments~\cite{fu2023dreamsim}.

\subsection{Conditional Diffusion Models}

Recently, conditional generative models based on diffusion models have seen significant development. The classifier-free guidance method has demonstrated strong controllable generative capabilities through supervised learning~\cite{ho2022classifier}. 
These methods dominate the tasks of text-to-image generation. From the widely used Stable Diffusion~\cite{rombach2022high,podell2023sdxl,sauer2023adversarial} to subsequent methods like ControlNet~\cite{zhang2023adding} and IP-Adapter~\cite{ye2023ip}, they have added more controllable conditions to conditional generation, such as edge maps, segmentation maps, depth maps~\cite{rombach2022high,ramesh2022hierarchical,meng2021sdedit,zhang2023adding,yu2023freedom,bansal2023universal}. 
Additionally, conditional generative models based on human feedback and preferences have also shown potential. Works like FABRIC and DPOK use real-time similarity tasks to guide models to generate images meeting human needs, proving that human feedback can provide more nuanced control over conditional generative models~\cite{von2023fabric,fan2023dpok,tang2023zeroth}. Methods like Pick-a-pic and Human Preference Score generate images aligned with human aesthetics~\cite{kirstain2023pick,wu2023human}. 
These advancements demonstrate that, in comparison to standard text-to-image frameworks, conditional generative models possess significant potential for more closely meeting human needs and preferences.

%%%%%%%%%%%%
\section{Discussion}

We proposed the CoCoG model, capable of predicting human visual similarity judgment behavior and learning human conceptual embeddings for visual objects. It can also efficiently and controllably generate visual objects in line with human cognition (Fig~\ref{fig:concept encoder}), manipulating human similarity judgment behavior and studying causal mechanisms in human cognition (Fig~\ref{fig:intervene}).

\paragraph{Contributions to AI} Our approach bridges generative models and human visual cognition. Through the concept encoding model, we align DNNs with human visual concept representations, simulating human processing and responses to visual objects more precisely, with potential to enhance AI's visual understanding capabilities; through controllable diffusion generation based on concept embeddings, we make conditional generative models more closely linked to human cognition and can manipulate human behavior through generated stimuli, promising to improve control and safety in AI-human interactions.

\paragraph{Contributions to cognitive science} Our approach significantly expands research on human visual cognition in cognitive science. With the concept encoding model, we achieved interpretable concept encoding for visual objects (Figure~\ref{fig:concept encoder}b); with controllable diffusion generation based on concept embeddings, we can generate a rich variety of natural stimuli to control human similarity judgment behaviors (Figure~\cite{roads2021enriching,fu2023dreamsim}). By combining advanced AI models with cognitive science research methods, we greatly enhance the efficiency and breadth of visual cognition research. Additionally, this method may reveal causal mechanisms in human visual cognition, offering new perspectives for understanding human cognitive processes.

\paragraph{Future directions} In the future, we will extend the paradigm of human visual cognition research to the study of AI representational spaces, which would help align AI with humans and provide new insights for understanding AI's cognition. Also, we recommend to bring optimal experimental design~\cite{rainforth2023modern,roads2021enriching} into human experiments. It will largely improve the efficiency of human behavioral data collection, facilitate model learning,  optimizing experimental paradigms with broader applications in cognitive science and AI.

%%%%%%%%%%%%

\section*{Ethical Statement}
The human behavioral data in this study is from THINGS public dataset. No animal or human experiments are involved. %The authors declare no competing interests. 

\section*{Acknowledgments}
This work is supported by the National Key R\&D Program of China (2021YFF1200804), Shenzhen Science and Technology Innovation Committee (20200925155957004, KCXFZ20201221173400001, SGDX2020110309280100), Guangdong Provincial Climbing Plan (pdjh2024c21310).

\section*{Appendix}

\appendix

\section{Concept encoder}

\subsection{Objective functions}

Our concept encoder is designed to learn human concept representations of visual objects by fitting human similarity judgment behavior. In line with prior studies~\cite{muttenthaler2022human}, we employ two strategies to train the concept encoder.

\paragraph{Fitting human behavior}

Given an image similarity \( S \) and a triplet \( \{c_i, c_j, c_k\} \) with \( c = g_{\theta}(f(x)) \) and \( S_{ij} = \langle c_i, c_j \rangle \), we use the softmax of object similarity to model the probability that \( \{a, b\} \in \{i, j, k\} \) is the most similar pair:

\begin{equation}
p(\{a, b\} | \{i, j, k\}) := \frac{\exp(S_{ab})}{\exp(S_{ij}) + \exp(S_{ik}) + \exp(S_{jk})}
\end{equation}

Our model aims to maximize the log-likelihood of the odd-one-out judgments. Diverging from previous work, we introduce an \( l_1 \) regularization term in the concept \( c \) to maintain the sparsity of concept embeddings:

\begin{equation}
\underset{\theta}{\text{arg min}} \ -\frac{1}{n} \sum_{s=1}^{n} \log p(\{a, b\} | \{i, j, k\}) + \lambda \| c \|
\end{equation}

\paragraph{Fitting VICE embeddings}

To facilitate comparisons with current methodologies, our model is also capable of fitting existing concept embeddings. As an example, we use VICE~\cite{muttenthaler2022vice} embeddings:

\begin{equation}
\underset{\theta}{\text{arg min}} \sum_{i=1}^{m} \| \bar{c}_m - g_{\theta}(f(x_m)) \| + \lambda \| c_m \|
\end{equation}

where \( \bar{c}_m \) represents the VICE dimensions for image \( m \), and \( c_m = g_{\theta}(f(x_m)) \) denotes the concept embeddings predicted by our model for image \( m \).

\subsection{Model comparison}

In Table~\ref{tab:modelcomparation}, we compare various network architectures in terms of their ability to predict behavioral outcomes. As shown, we have evaluated different combinations of linear projection (LinProj) and multi-layer perceptron (MLP) with embedding-based (Emb) and behavior-based (Beh) objective functions. The results are benchmarked against a baseline (Chance) and an upper bound (noise ceiling) to provide context for their performance.

From these comparisons, we can observe that the MLP models, especially when combined with behavior-based objectives, tend to offer superior performance, as indicated by their higher accuracy in the test set. This suggests that the MLP architecture, particularly when aligned with behavior-oriented training, is more effective in capturing the nuances of human conceptual representations. But for comparison with VICE, we use linear projection mbedding-based objective functions for subsequent concept decoder.

\begin{table}[!h]
\renewcommand{\arraystretch}{1.1} 
  \caption{Comparison of different networks architectures for behavioral prediction. The accuracy in the test set for each model is listed. CoCoG variants of the concept encoder are compared to a baseline (Chance) and an upper bound (noise ceiling). Emb and Beh mean different objective functions. }
  \label{tab:modelcomparation}
  \centering
  \begin{threeparttable}
  \begin{tabular}{lc}
    \toprule
    & \multicolumn{1}{c}{Behavioral prediction}\\
    \midrule
    \textcolor{gray}{Chance}\tnote{*}    &   \textcolor{gray}{33.33}\%   \\
    \midrule
    LinProj + Emb &  61.90\%   \\
    MLP + Emb   &  63.02\%   \\
    LinProj + Beh &  62.53\%   \\
    MLP + Beh   &  64.07\%   \\
    \midrule
    \textcolor{gray}{noise ceiling}\tnote{*}   &  \textcolor{gray}{ 67.22\%}      \\
    \bottomrule
  \end{tabular}
    \end{threeparttable}
\end{table}

\subsection{Additional results for model validation}

\begin{figure}[!h]
\centering
\includegraphics[width=7.5cm]{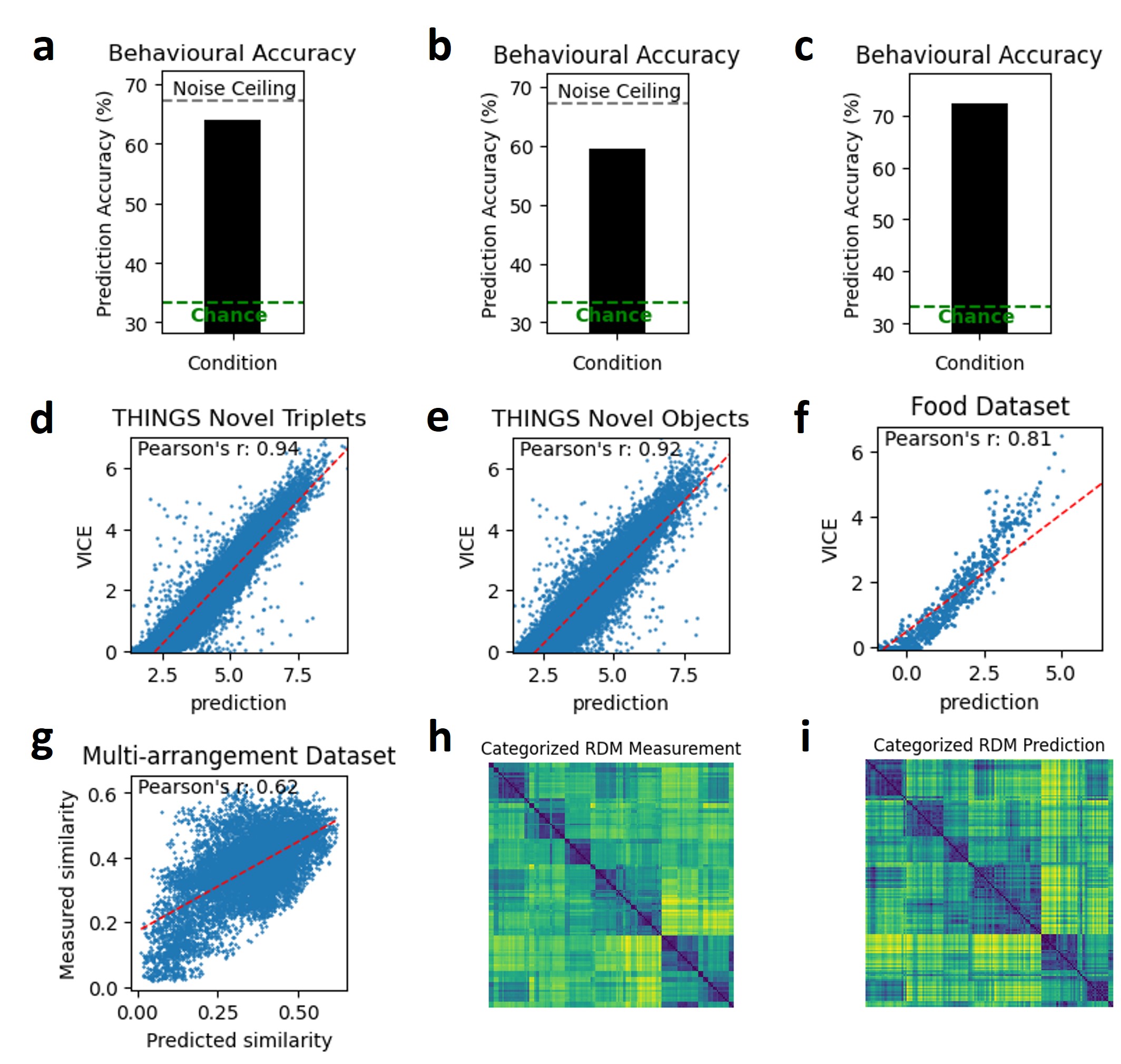}
\caption{Model generalization performance within a dataset, across datasets, and across tasks.}
\label{fig:generalization}
\end{figure}

\paragraph{Generalization across objects, datasets, and tasks}
Figure~\ref{fig:generalization} demonstrates our model's performance within and across datasets, as well as its adaptability to different task paradigms.
In the context of the THINGS dataset~\cite{hebart2023things}, our model exhibits strong generalization capabilities, as evidenced by its accuracy in predicting novel odd-one-out decisions in both the validation set (Novel Triplets) and a subset of objects (Novel Objects), achieving accuracies of 64.07\% and 59.55\%, respectively. The object similarity computed by our model's embeddings correlates highly with the VICE embeddings, showcasing Pearson correlation coefficients of 0.94 and 0.92 for the Novel Triplets and Novel Objects experiments, respectively. These results underscore the robustness of our model in handling within-dataset variations.

When focusing on the cross-dataset scenario with the Food Dataset~\cite{carrington2024naturalistic}, our model continues to maintain high performance standards. Here, it achieves a prediction accuracy of 72.41\% on the validation set after training on the respective train set. The object similarity calculated by our model in this novel food dataset exhibits a significant Pearson correlation with VICE embeddings, indicated by a coefficient of 0.81. This outcome demonstrates the model's capacity to adapt and perform effectively across different content domains.

Additionally, we explored the cross-task generalization ability of our model through the Multi-arrangement task paradigm~\cite{king2019similarity}. When trained on this new task paradigm, the object similarity calculated by our model showed a notable correlation with the experimental data, indicated by a Pearson coefficient of 0.62. The alignment of model-predicted and experimentally measured representation distance matrices, particularly concerning object classification, further validates our model's efficacy.

\begin{figure}[!h]
\centering
\includegraphics[width=7.5cm]{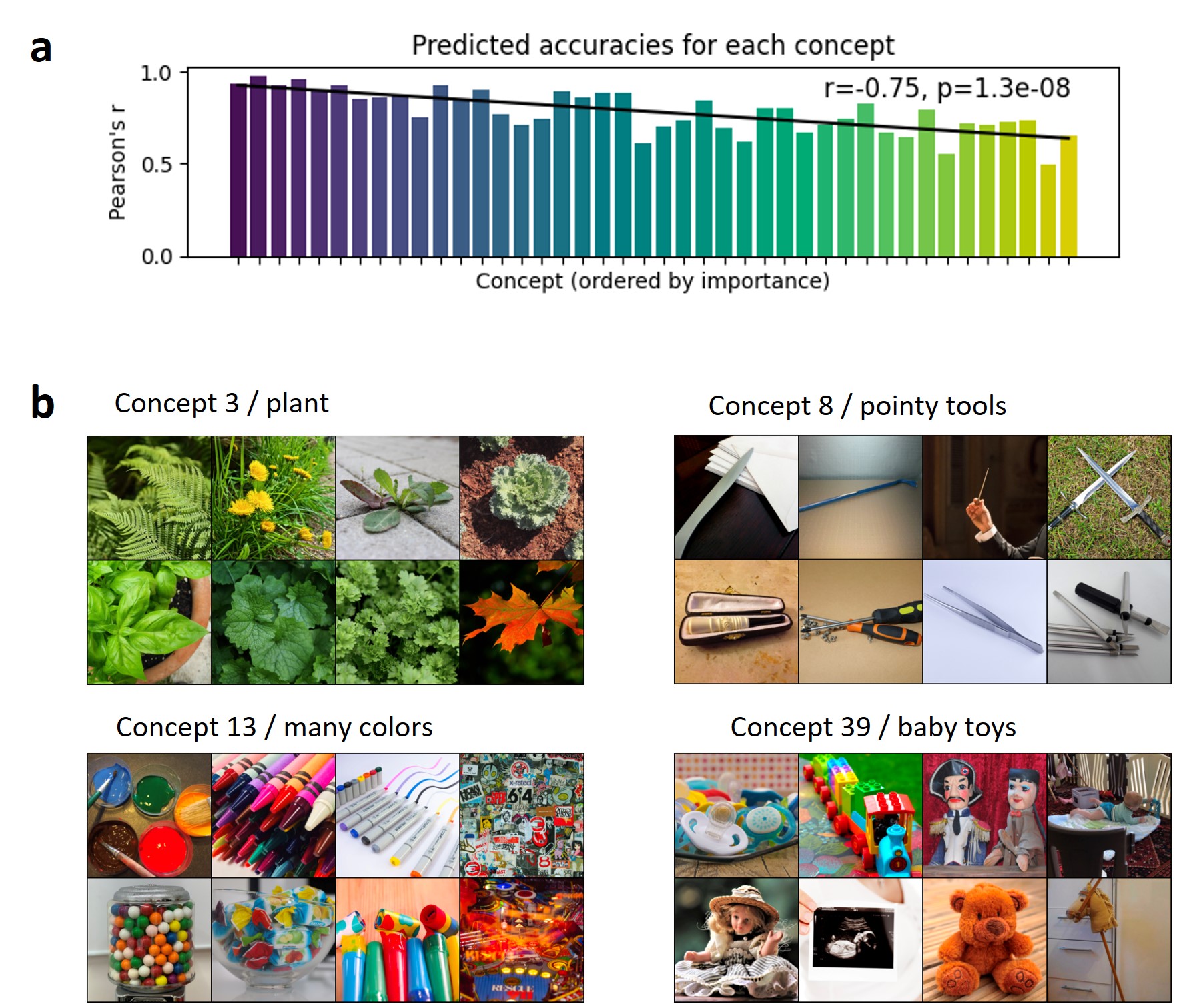}
\caption{Concept prediction by concept encoder. (a) Predicted accuracies for each concept of VICE; (b) images with high activation in selected concepts.}
\label{fig:concept}
\end{figure}

\paragraph{Concept encoder performance}

Figure~\ref{fig:concept} showcases the proficiency of our concept encoder in extracting and representing various concepts. Our analysis reveals a significant negative correlation between the importance of a concept and its predictability, quantified by Pearson's r. Specifically, the correlation coefficient for this relationship stands at -0.75 (p = 1.3e-8), suggesting that concepts deemed more important tend to be more predictable, which aligns with our initial expectations. For this evaluation, we utilized concept embeddings from previous research as a benchmark for comparison.

Complementing this quantitative analysis, the visual representations provided in the figure illustrate our model's effectiveness in activating multiple conceptual dimensions. These example images clearly demonstrate the model's nuanced capability to recognize and encode a diverse range of concepts, thereby reaffirming its applicability and effectiveness in complex conceptual understanding tasks.

\begin{figure}[!h]
\centering
\includegraphics[width=7.5cm]{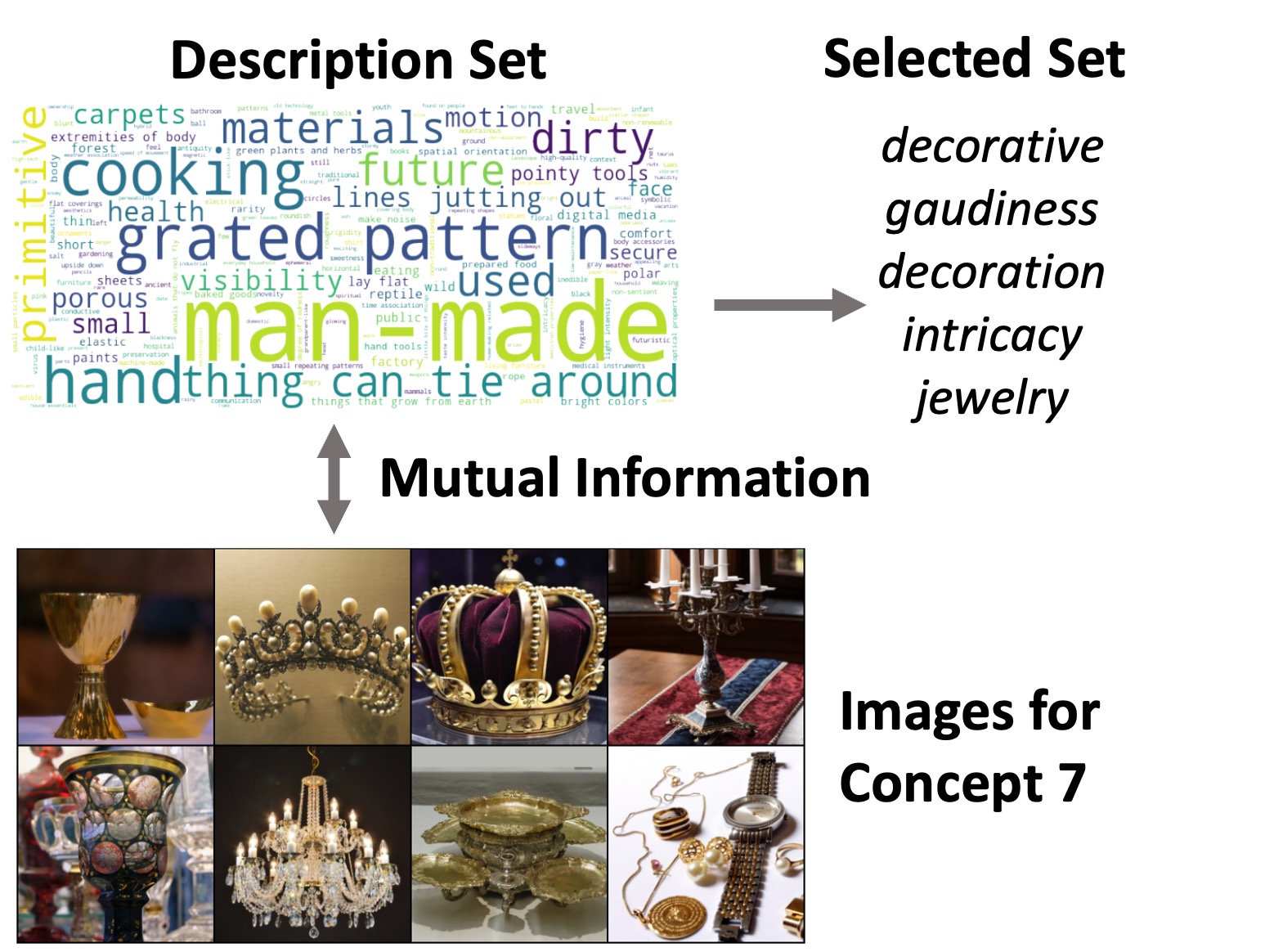}
\caption{Model for concept description selection.}
\label{fig:language model}
\end{figure}

\subsection{Description of concepts}

Inspired by the CLIP-Dissect model~\cite{oikarinen2022clip}, we adopt a mutual information maximization method to select descriptions for each concept dimension. As depicted in Figure~\ref{fig:language model}, we compiled a descriptive word dataset \( T \) containing 23,153 words. For a given concept \( C_k \), we calculate the mutual information between the descriptions and the maximum activation visual objects \( B_k \) of the concept dimension, selecting words that yield the highest mutual information. We employ the SoftWPMI equation, as utilized by CLIP-Dissect:

\begin{equation}
\text{sim}(t_m, q_k; P) = \log \mathbb{E}[p(t_m | B_k)] - \lambda \log p(t_m)
\end{equation}

\begin{table*}[!h]
\centering
\renewcommand{\arraystretch}{1.2} 
\begin{tabular}{c|l}
\hline
\textbf{Index} & \textbf{Language Description} \\ \hline
1 & home tools, metal tools, hand tools, metallic tools, tools \\ \hline
2 & baked food, prepared food, food, made dish, entrees \\ \hline
3 & plant, green plants, green plants and herbs, plants, common in outdoor \\ \hline
4 & ground animals, small animals, land animals, mammal, animal type \\ \hline
5 & household furniture, furniture, main component of room, affordability, living furniture \\ \hline
6 & cotton clothing, clothing, clothes, things to wear, unisex \\ \hline
7 & decorative, gaudiness, decoration, intricacy, jewelry \\ \hline
8 & outdoor objects, building materials, ground, foundation, protruding \\ \hline
9 & pointy tools, hand tools, sharp tools, metal tools, metallic tools \\ \hline
10 & wood, wood-colored, made of wood, furniture, building materials \\ \hline
\end{tabular}
\caption{Five language descriptions found by the model that best match the first ten concepts.}
\label{tab:language_descriptions}
\end{table*}

\section{Concept decoder}

In this section, we provide a concise overview of the conditional diffusion model framework used in our concept decoder, following the presentation of continuous-time diffusion models in~\cite{song2020score,karras2022elucidating}.

\paragraph{Diffusion models} Diffusion Models (DMs) engage in a generative process by transforming high-variance Gaussian noise into structured data representations. This transformation is achieved by gradually reducing noise levels across a sequence of steps. Specifically, we begin with a high-variance Gaussian noise \( x_M \sim \mathcal{N}(0, \sigma_{\text{max}}^2) \) and systematically denoise it through a series of steps to obtain \( x_t \sim p(x_t; t) \), where \( \sigma_t < \sigma_{t+1} \) and \( \sigma_M = \sigma_{\text{max}} \). For a well-calibrated DM, and with \( \sigma_0 = 0 \), the final \( x_0 \) aligns with the original data distribution.

\paragraph{Sampling process} The sampling in DMs is implemented by numerically simulating a Probability Flow ordinary differential equation (ODE) or a stochastic differential equation (SDE). The ODE is represented as:
\begin{equation}
dx = -\dot{\sigma}(t)\sigma(t)\nabla_x \log p(x; t) dt,
\end{equation}
where \( \nabla_x \log p(x; t) \) is the score function, and \( \sigma(t) \) is a pre-defined schedule with its time derivative \( \dot{\sigma}(t) \). The SDE variant includes a Langevin diffusion component and is expressed as:
\begin{equation}
\begin{split}
dx = &-\dot{\sigma}(t)\sigma(t)\nabla_x \log p(x; t) dt \\
& - \beta(t)\sigma^2(t)\nabla_x \log p(x; t) dt \\
& + \sqrt{2\beta(t)}\sigma(t) d\omega_t,
\end{split}
\end{equation}
where \( d\omega_t \) is the standard Wiener process.

\paragraph{Training of DMs} The core of DM training is to learn a model \( s_{\theta}(x; t) \) for the score function. This is typically achieved through denoising score matching (DSM), where \( \epsilon_{\theta} \) is a learnable denoiser. The training process can be formulated as:
\begin{equation}
\mathbb{E}_{(x_0, c) \sim p_{\text{data}}(x_0, c), (n_t, t) \sim p(n_t, t)} \left[ \| \epsilon_{\theta}(x_0 + n_t; t, c) - x_0 \|_2^2 \right],
\end{equation}
where \(n_t\) is Gaussian noise with variance \(\sigma_t^2\), and \( c \) represents a condition.

\subsection{Stage I - prior diffusion}

The training of the prior diffusion stage is a critical step in our concept decoder, employing the classifier-free guidance technique in conjunction with data pairs of CLIP embeddings and concept embeddings \((h_i, c_i)\).
Drawing on principles from advanced generative models, our prior diffusion process is conditioned on the concept embedding \( c \) to effectively learn the distribution of CLIP embeddings \( p(h|c) \). The CLIP embedding \( h \) obtained in this stage serves as the prior for the subsequent stage. Our model architecture includes a lightweight U-Net, denoted as \( \epsilon_{prior}(h_t, t, c) \), where \( h_t \) represents the noisy CLIP embedding at the diffusion time step \( t \). Training pairs consisting of CLIP and concept embeddings are extracted from the ImageNet dataset, which have over 1 million images. These pairs are then used to train the Prior Diffusion model. The model is proficiently trained using the classifier-free guidance method, which ensures a balance between fidelity to the conditioning signal and the generative diversity of the outputs.

\paragraph{Classifier-free guidance technique} The Classifier-Free Guidance method plays a pivotal role in directing the iterative sampling of a Diffusion Model (DM) in response to a given condition, specifically a concept \( c \). This technique operates by harmonizing the outputs from both a conditional model and an unconditional model. The formulation of the combined model, \( \epsilon^w_{prior}(h; t, c) \), is given as:
\begin{equation}
\epsilon^w_{prior}(h; t, c) = (1 + w)\epsilon_{prior}(h; t, c) - w\epsilon_{prior}(h; t),
\end{equation}
where \( w \geq 0 \) indicates the {\it guidance scale}. This approach enables simultaneous training of the unconditional model alongside the conditional model within a single network architecture. It is accomplished by intermittently replacing the concept \( c \) with a null vector, as per Equation (3), at a fixed proportion of the time, like 10\%. The chief application of this method is to refine the quality of the samples generated by DMs, balancing it against the diversity of the outputs.

\subsection{Stage II - CLIP guidance generation}

In Stage II of our concept decoder, the CLIP embedding \( h \) obtained from the prior diffusion training serves as the basis for generating visual objects \( x \) conditioned on \( h \). For this purpose, we employ the integration of pre-trained models, namely SDXL and IP-Adapter~\cite{podell2023sdxl,ye2023ip}, to achieve efficient and high-quality image generation.

The backbone of our image generation process is the SDXL model, known for its robustness in text-to-image diffusion processes. By incorporating the IP-Adapter, which introduces dual cross-attention modules, the CLIP embedding \( h \) is effectively used as a conditional input, thereby guiding the denoising process in the U-Net architecture. The combined model for this process is denoted as \( \epsilon_{SD}(z_t, t, h) \), where \( z_t \) represents the noisy latents from SDXL's Variational Autoencoder (VAE).

\paragraph{Advantages of pre-Trained model integration} Utilizing these pre-trained models allows us to leverage their inherent capabilities without the need for additional modifications. This approach not only simplifies the implementation process but also ensures the retention of the quality attributes inherent in these models. The integration of concept embedding guidance with these pre-trained models facilitates the generation of visual objects that are closely aligned with the concept information encoded in \( h \).

\paragraph{SDXL-turbo for enhanced speed} To further enhance the efficiency of our model, we also explore the use of SDXL-Turbo~\cite{sauer2023adversarial}, a distilled version of SDXL, optimized for real-time synthesis. This model is particularly advantageous in scenarios where rapid generation of high-fidelity images is required.

\paragraph{IP-Adapter} The IP-Adapter, with its relatively lightweight architecture, has demonstrated its effectiveness in adding image prompt capability to pre-trained text-to-image models. Its ability to work in tandem with text prompts for multimodal image generation broadens the scope of applications for our concept decoder.

\subsection{Performance Metrics}

\paragraph{Similarity}
We define similarity using the dot product as mentioned in the Objective Functions section. Specifically, the similarity \( S_{ij} \) between the target concept embedding and the concept embedding extracted from the generated image is calculated as:
\begin{equation}
S_{ij} = \langle c_i, c_j \rangle
\end{equation}
where \( c_i \) and \( c_j \) represent the respective concept embeddings.

\paragraph{Diversity}
Inspired by works in the literature~\cite{von2023fabric,corso2023particle}, we describe image diversity for a single concept embedding using In-batch CLIP similarity. We use this approach to assess the variety of images generated from the same concept embedding. In-batch CLIP similarity quantifies how different the generated images are from one another within the same batch, reflecting the diversity of outputs for a given concept embedding. The In-batch Image Diversity \( d \) for a batch of images \( x_1, \ldots, x_n \) is defined as follows:
\begin{equation}
d(x_1, \ldots, x_n) = 1 - \frac{2}{n(n - 1)} \sum_{i=2}^{n} \sum_{j=1}^{i-1} \text{CLIP}(x_i, x_j)
\end{equation}
where \(\frac{n(n-1)}{2}\) is the number of elements in the upper triangular cosine similarity matrix.

\bibliographystyle{named}
\bibliography{main}

\end{document}